\documentclass[12pt]{article}

\usepackage{amsmath}
\usepackage{epsfig}
\usepackage{amsfonts}

\newtheorem{theorem}{Theorem}
\newtheorem{lemma}{Lemma}

\newcommand{\T}{\ensuremath{\mathcal{T}}}
\newcommand{\MAX}{{\sc Max}}
\newcommand{\MIN}{{\sc Min}}
\newcommand{\AND}{{\sc And}}
\newcommand{\OR}{{\sc Or}}

\newcommand{\MINMAX}{\MIN-\MAX}
\newcommand{\ANDOR}{\AND-\OR}

\newcommand{\val}{\mbox{\rm value}\hspace*{0.1mm}}

\newcommand{\floor}[1]{\ensuremath{\lfloor{#1}\rfloor}}
\renewcommand{\epsilon}{\varepsilon}

\newcommand{\half}{\frac{1}{2}}

\begin{document}
\title{\bf Quantum Algorithms for \\ Evaluating M{\large IN}-M{\large AX} Trees}

\author{Richard Cleve%
\thanks{David R.~Cheriton School of Computer Science and Institute for Quantum Computing,
University of Waterloo.}%
\ \thanks{Perimeter Institute, Waterloo, Ontario, Canada.}
\and
Dmitry Gavinsky$^{\ast}$
\and
David L.~Yonge-Mallo$^{\ast}$}

\date{31 Oct 2007}

\maketitle

%


\vspace{-2ex}
\begin{abstract}
\noindent
We present a bounded-error quantum algorithm for evaluating \MINMAX{} 
trees with $N^{\half+o(1)}$ queries, where $N$ is the size of the tree 
and where the allowable queries are comparisons of the form $[x_j < x_k]$.
This is close to tight, since there is a known quantum lower bound of 
$\Omega(N^{\half})$.
\end{abstract}

\maketitle 


\noindent
A \MINMAX{} tree is a tree whose internal nodes are \textit{minimum} 
and \textit{maximum} gates, at alternating levels, and whose 
leaves are values from some underlying ordered set.
The size $N$ of such a tree the number of its leaves, whose values are 
referred to as $x_1, \ldots, x_N$.
The value of a \MINMAX{} tree is the value of its root, a function 
of $x_1, \ldots, x_N$.
In the \textit{input value} query model, queries explicitly access the 
values of the leaves.
In the \textit{comparison} query model, the values of $x_1, \ldots, x_N$ 
are not directly accessible; rather, queries are comparisons of the form 
$[x_j < x_k]$.
In this latter model, the appropriate output is any $j \in \{1,\ldots,N\}$ 
such that $x_j$ is the value of the tree.

Note that, when the ordered set is $\{0,1\}$, a \MINMAX{} tree reduces 
to an \ANDOR{} tree.
This implies that Barnum and Saks's lower bound of 
$\Omega(N^{\half})$~\cite{barnum2004:read-once} for the quantum query 
complexity of \ANDOR{} trees applies to \MINMAX{} trees.

Recent results initiated by Farhi \textit{et al.} have shown that quantum 
algorithms can evaluate
all \ANDOR{} trees with order $N^{\half+o(1)}$ queries
\cite{farhi2007:quantum,childs2007:quantum,childs2007:every_NAND_formula,ambainis2007:nand}.
We show that these results carry over to \MINMAX{} trees in 
both the input value model and the comparison model.

Let $W(N)$ be the query complexity for \ANDOR{} trees of size $N$.
We show that \MINMAX{} trees can be evaluated with 
$O(W(N)\log(N))$ queries in both the input value model and 
the comparison model.
Our algorithm combines the results on \ANDOR{} trees in 
Refs.~\cite{ambainis2007:nand,childs2007:every_NAND_formula} 
with the lemma below and Grover's search 
algorithm~\cite{grover1996:search}.

\begin{lemma}\label{lem:AND-OR_threshold}
Let $\T$ be a \MINMAX{} tree with inputs
$x_{1}, x_{2}, \ldots, x_{N}$.
Let $\T^v$ be an \ANDOR{} tree with identical
structure to $\T$, but with \AND{} and \OR{}
gates in place of \MIN{} and \MAX{} gates (respectively),
and with the $k^{\mbox{\scriptsize th}}$ input assigned
to $1$ if and only if $x_{k} \geq v$.
Then $\val(\T^v) = 1$ if and only if $\val(\T) \geq v$.
\end{lemma}

Lemma~\ref{lem:AND-OR_threshold} is easy to prove by induction.
It implies that, if the underlying ordered 
set is a numerical range of size $N^{O(1)}$, then the tree can be 
evaluated in $\log(N)$ stages by a simple binary search.
Each stage can be implemented with $O(W(N)\log\log(N))$ queries, 
which reflects the cost of evaluating an \ANDOR{} tree amplified so that 
its error probability is $O(1/\log(N))$.
The result is an $O(W(N)\log(N)\log\log(N))$ query algorithm.

A complication arises in performing such a binary search in the comparison 
model, where it is not possible to directly compute the midpoint of 
an interval like $[x_j, x_k]$.
Problems can also arise in the input value model when the numerical range is 
too large: the binary search may not converge in a logarithmic number of steps.
For this reason, we avoid the standard binary search approach where a midpoint 
is chosen as a pivot. 
Instead, we take a \textit{random} input value among those that that lie 
within a current interval as our pivot.
What is noteworthy about this simple approach is that \textit{it does not 
work efficiently in the classical case}: given an interval $[x_j, x_k]$, 
finding an interior point is as hard as searching, which can cost $\Omega(N)$ 
queries to do even once \cite{bennett1997:lowerbound}.
In the setting of \textit{quantum} algorithms, we can utilize Grover's search 
algorithm~\cite{grover1996:search,boyer1998:tight_bounds} which costs $O(\sqrt{N})$.

As an aside, we note that there is a classical reduction from \MINMAX{} 
trees to \ANDOR{} trees that yields an $O(N^{0.753})$ query algorithm for  
balanced \MINMAX{} trees~\cite{saks1986:game_trees}.
We can use that reduction with an $N^{\half}$ query \textit{quantum} 
algorithm for balanced \ANDOR{} trees; however, the resulting algorithm for
\MINMAX{} costs $\Omega(N^{0.58})$.
Our alternate approach yields exponent $\half + o(1)$ and is not restricted 
to balanced trees.


What follows is a description of our algorithm with the analysis of its error.
For convenience, let $\bot$ and $\top$ be such that $x_\bot<x_j$ and $x_\top>x_j$ 
for any $j \in \{1,\ldots,N\}$ and let $c$ be a constant.

~\\
\noindent {\sc Quantum \MINMAX{} tree evaluation}
\begin{enumerate}

\item

    Let $\gamma \leftarrow \bot$ and $\delta \leftarrow \top$, and
    initialize the stack.

\item Repeat the following steps for $c \log(N)$ iterations, then go to Step 3:

    \begin{enumerate}

    \item Find a random pivot:

        Call the quantum search subroutine to find a random
        pivot index $j$ with $x_{\gamma} < x_{j} < x_{\delta}$.
        If no value is found, 
        go to Step~2(c).

    \item Refine the search:

        Call the \ANDOR{} tree subroutine to check if
        $\val(\T) < x_{j}$. If so, let
        $\delta \leftarrow j$; otherwise, let
        $\gamma \leftarrow j$.

    \item Backtrack if out of range:

        Call the \ANDOR{} subroutine to check if
        $x_{\gamma} \leq \val(\T) < x_{\delta}$.
        If so, push $(\gamma, \delta)$ onto the stack.
        Otherwise, pop $(\gamma, \delta)$ off the stack.
        (If the stack is empty, let 
        $\gamma \leftarrow \bot$ and $\delta \leftarrow \top$.)

    \end{enumerate}

\item
    Return $\gamma$ as an index corresponding to the value of the \MINMAX{} tree.

\end{enumerate}

Clearly, the algorithm makes $O(W(N)\log(N))$ queries.
We claim the following.

\begin{theorem}\label{thm:correctness}
The algorithm returns the value of the \MINMAX{} tree with
probability at least $\frac{2}{3}$.
\end{theorem}

To prove Theorem~\ref{thm:correctness}, we must consider
the progress made by the random choices of pivots as well
as the error probabilities of the subroutines for \ANDOR{}
and the searches (each errs with constant probability).

To begin with, assume that the subroutines for \ANDOR{}
and search never err (thus, $x_{\gamma} \leq \val(\T) < x_{\delta}$
at all times).
Under this assumption, the progress of the algorithm is
determined by how quickly the subinterval converges.
Once no value in Step~2(a) is found 
the algorithm has \textit{converged} 
(with $x_{\gamma} = \val(\T)$)
and can go to Step~3 and terminate (however
it is harmless to perform more iterations before doing this).

Let $C(m)$ denote the expected number of iterations of the algorithm
until it converges, assuming that $m$ of its inputs are
within its current range.

Then, for $m > 1$, $C(m)$ satisfies the recurrence
\begin{equation}\label{eqn:recurrence}
C(m) \leq \frac{2}{m} \left( \sum_{k = \floor{m/2}}^{m-1} C(k) \right) + 1.
\end{equation}
This can be seen by assuming that the pivot is uniformly selected among all
$m$ possible positions within the subinterval and that $\val(\T)$ always
lies in the larger side of the pivot.
It is straightforward to verify that the recurrence implies
$C(m) \in O(\log(m))$.
Therefore, the expected number of iterations of Step~2
made by the algorithm before $x_{\gamma} = \val(\T)$, under the
assumption that the subroutines never
err, is $O(\log(N))$.
By the Markov bound, $O(\log(N))$ iterations suffice to obtain
error probability less than any particular constant.

We now consider the fact that the subroutines for \ANDOR{} and
searching can fail.
First, note that, by incurring a multiplicative factor of
only $O(\log\log(N))$, each call to the \ANDOR{} and search 
algorithm can be amplified so that its error probability is
$O(1/\log(N))$.
This results in an $O(W(N)\log(N)\log\log(N))$ algorithm
for \MINMAX.

These amplification costs are not necessary in our algorithm,
since it can cope with a constant fraction of errors in
subroutine calls.  To see why this is so, let $\varepsilon$
be the probability that one or more subroutines err during
one iteration of Step~2 of the algorithm.  
The algorithm begins some $O(\log(N))$ steps away 
from reaching a \textit{good} state\,---\,of the form $(\gamma',\delta)$ 
such that $x_{\gamma'} = \val(\T)$.
Before reaching a good state, an ``incorrect'' step for the 
algorithm places $\val(\T)$ outside the search interval, and a
``correct'' step either narrows the search interval or
backtracks from a previous error.  After reaching 
a good state, a ``correct'' step pushes a pair of the form 
$(\gamma',\delta)$ onto the stack and an ``incorrect'' step pops it off.
In each iteration, the algorithm takes a correct
step with probability at least $1 - \varepsilon$ and
an incorrect step with probability at most $\varepsilon$.
Therefore, with all but exponentially small probability, the number 
of correct steps minus the number of incorrect ones after $c \log(N)$ 
iterations is at least $\frac{c}{2}\log(N)$.
For suitably large $c$ this means that, with constant probability, 
when the algorithm terminates, $x_{\gamma} = \val(\T)$ (typically with many 
copies of pairs of the form $(\gamma',\delta)$ on the top of its stack).


Finally, we note that, in game-playing contexts, it is useful to 
determine optimal moves.
This corresponds to finding the subtree of a \MINMAX{} tree that 
attains its value.
If the leaf values $x_{1},\ldots,x_{N}$ are distinct, this is easily 
deduced from $\val(\T)$.
Otherwise, one can use a slightly modified version of the minimum/maximum 
finding algorithm in Ref.~\cite{durr1996:quantum_finding_minimum} to 
find the appropriate subtree. 

\section*{Acknowledgements}
\noindent
We would like to thank Peter van Beek, Peter H\o yer and Pascal Poupart 
for helpful discussions.
This research was supported in part by Canada's NSERC, CIAR, MITACS, QuantumWorks, 
and the U.S.~ARO/DTO.



\end{document}